\documentclass[12pt]{iopart}
\usepackage{graphicx}
\newcommand{\ket}[1]{|{#1}\rangle}
\newcommand{\bra}[1]{\langle{#1}|}
\newcommand{\ketbra}[2]{|{#1}\rangle\langle{#2}|}

\begin{document}
\title{Decoherence at constant excitation}
\author{J. M. Torres$^1$, E. Sadurn\'i$^2$, 
T. H. Seligman$^{1,3}$ }
\address{$^1$Instituto de Ciencias F\'isicas,
Universidad Nacional Aut\'onoma de M\'exico,
C.P. 62210 Cuernavaca, Morelos, M\'exico}
\address{$^2$Universit\"at Ulm, Institut f\"ur 
Quantenphysik, D-89069 Ulm, Germany}
\address{$^3$Centro Internacional de Ciencias,
C.P. 62210 Cuernavaca, Morelos, M\'exico}
\ead{mau@fis.unam.mx}

\begin{abstract}
We present a simple exactly solvable extension of
of the Jaynes-Cummings model by adding 
dissipation. This is done such that the total
number of excitations is conserved.
The Liouville operator in the 
resulting master equation can
be reduced to blocks of $4\times 4$
matrices.
\end{abstract}

 
 
\maketitle
\section{Introduction}

In the field of quantum optics it is typical to deal with open systems in which
some degrees of freedom are treated as an environment.
The simplest example is a two level atom
in thermal equilibrium with a continuum of modes  \cite{carmichael}.
This accounts for an effective spontaneous decay from
an excited state to the ground state. To deal with this
kind of problems, one usually works under a Markovian
assumption (environment without memory) and as a result
one can work with a master equation in the Lindblad form \cite{lindblad},
which also follows from the pioneering work of Kossakowski
and collaborators \cite{kossakowski, gorini}.

Solutions to this kind of equation, except for a two level
atom, involve complicated expressions due to the extended 
space in which one is working. The master equation can be written
as the action of a super-operator or Liouville operator 
${\mathcal L}$ on a density matrix $\rho$. If one has an
$n\times n$ density matrix,  the super-operator will act on a $n\times n$ dimensional space.

In this work we present a new form for a Jaynes
Cummings model with dissipation. The aim here is simply exact solvability with 
Kraus operators not considered previously.
 We show how to construct a master equation whose Liouvillian can
be diagonalized in blocks of $4\times  4$ matrices and thus produce analytical solutions.
Yet we will see that in the same way more complicated problems, such as spins larger than 
$1/2$ can be brought to finite matrix block form and therefore at least allow numerical 
treatment to arbitrary accuracy.

Indeed in two recent papers we displayed models solvable in closed form for two different atoms in a cavity and for Dirac-Moshinsky oscillators coupled to an isospin field \cite{torres, emerson} . These models were based on the fact, that Hamiltonians could be constructed, such that the total number of excitations was an additional conserved quantum number. In the basis where this quantity was diagonal the Hamiltonian broke into blocks no larger than 
$4\times 4$.

Here we make an analogue construction for the master equation in the Lindblad 
form for the Jaynes Cummings model in rotating wave approximation using Kraus operators that preserve the total number of excitations.
This Hamiltonian reduces to $2\times 2$ blocks and the corresponding super-operator to 
$4\times 4$ blocks again thus
guaranteeing solvability.

In section \ref{sec:JC} we present a review of the Jaynes-Cummings model, which
will allow us to fix the notation to be used in section \ref{sec:master} where
we introduce the explicit master equation that describes a system with dissipation,
but at constant number of total excitations. This allows a spectral decomposition
of the Liouville operator in terms of $4\times 4$ matrices. In section 
\ref{sec:det0} we present explicit solutions to the eigenvalue problem when the
atom is in resonance with the cavity and discuss the dynamics of two particular
initial states. 

\section{The JC model}\label{sec:JC}
Consider the Jaynes-Cummings Hamiltonian in the  rotating wave approximation
\begin{equation}
H=\delta\sigma_z+g\left(a\sigma_++a^\dagger\sigma_-\right).
\end{equation}
One can identify an additional  conserved quantity, which
can be interpreted as the number of excitations
\begin{equation}
  I=a^\dagger a+\frac{1}{2}\left(\sigma_z+{\bf 1}\right).
\end{equation}
The basis in which $I$ is diagonal is given by the states
\begin{eqnarray}
  \ket{n-1,1},\quad
  \ket{n,0},
\end{eqnarray}
where $\ket{n-1,1}$ represents a state with $n-1$
photons in the field mode and an excited atom.
$\ket{n,0}$ accounts for $n$ photons and 
the atom in the ground state.
The action of $I$ on each of the previous states is given
by
\begin{equation}
  I\ket{n-j,j}=n\ket{n-j,j},\quad j=0,1.
\end{equation}
In this basis $H$ is block-diagonal, with
the blocks
\begin{equation}\label{eq:matjc}
  H_n=\left(
  \begin{array}{cc}
    \delta&g\sqrt{n}\\
    g\sqrt{n}&-\delta
  \end{array}
  \right),
\end{equation}
with the corresponding eigenenergies 
\begin{equation}
  \pm E_n=\pm\sqrt{\delta^2+g^2 n}.
\end{equation}
The eigenstates -the dressed states- of $H$ are given by
\begin{eqnarray}\label{eq:dress}
  \ket{\phi_n^+}&=
  \cos{\theta_n}\ket{n-1,1}
  +
  \sin{\theta_n}\ket{n,0}
  ,\nonumber\\
  \ket{\phi_n^-}&=
  -\sin{\theta_n}\ket{n-1,1}
  +\cos{\theta_n}\ket{n,0}
\end{eqnarray}
with
\begin{equation}
\theta_n=\arctan{\left(\sqrt{\frac{E_n-\delta}{E_n+\delta}}\right)}.
\end{equation}

\section{Master equation}\label{sec:master}
The Heisenberg equation for a density matrix 
in a closed system is given by  
\begin{equation}
  \dot\rho=-i\left[H,\rho\right]
\end{equation}
For an open system description 
one can consider a master equation 
in the Lindblad form which describes a non-unitary Markovian 
evolution \cite{lindblad}.  It can be written in terms of an arbitrary set of
Kraus operators $O_j$ as 
\begin{eqnarray}
  \dot\rho=&{\mathcal L}\rho
  \nonumber\\
  =&
  -i\left[H,\rho\right]
  -\sum_j
  \frac{\gamma_j}{2}\left(
  O_j^\dagger O_j\rho+\rho O_j^\dagger O_j
  -2 O_j\rho O_j^\dagger
  \right).
\end{eqnarray}
Where we have introduced the Liouville super-operator
$\mathcal L$ that acts on a density matrix $\rho$. $\mathcal L$ is
a linear operator and the generator of a completely positive dynamical 
semigroup \cite{gorini}.
Known solvable  models \cite{eiselt, briegel, daeubler} for this 
type of equation are, for instance, taking
one operator $O=a$ for a cavity with losses,
or $O=\sigma_z$ for an atom with spontaneous
decay.

In this work we explore the possibility of having other types of
operators which allow closed solution
of the master equation. 
The condition we
impose is that such operators commute with $I$,
this means that this dissipation will conserve the number
of excitations. 
There are many other operators that one could consider for instance  $I$ itself,
or $a^\dagger a$ and $\sigma_z$. Here we arbitrarily chose
the pair $O_1=a\sigma_+$ and  $O_2=a^\dagger\sigma_-$. With this
we can construct the following master equation 
which describes dissipative dynamics
\begin{eqnarray}\label{eq:master}
  \dot\rho&={\mathcal L}\rho=-i\left[H,\rho\right]
  \nonumber\\
  &-\frac{\gamma_0}{2}\left(
  a\sigma_+a^\dagger \sigma_-\rho+\rho a\sigma_+a^\dagger \sigma_-
  -2a^\dagger \sigma_-\rho a\sigma_+
  \right)
  \nonumber\\
  &-\frac{\gamma_1}{2}\left(
  a^\dagger\sigma_-a \sigma_+\rho+\rho a^\dagger\sigma_-a \sigma_+
  -2a \sigma_+\rho a^\dagger\sigma_-
  \right)
  .
\end{eqnarray}

As mentioned above, the operators we have chosen preserve the number of
excitations $I$. Then we find it convenient to work in  
the basis in which $I$ is diagonal to
to represent any density matrix as
\begin{equation}
  \rho=\sum_{n,m=0}^\infty\sum_{j,k=0}^1
  \rho_{n,m}^{j,k}
  \ketbra{n-j,j}{m-k,k}=
  \sum_{n,m}\rho_{n,m}.
\end{equation}
Here we have partitioned the density matrix into the $2\times 2$ matrices
$\rho_{n,m}$ with matrix elements $\rho_{n,m}^{j,k}$.
Note that for $n=0$ there is a single state as
$n-j\ge 0$, which in this case implies $j=0$.
Each $\rho_{n,m}$ has a definite number of left and right
excitations, which can be summarized by its commutation
relation with $I$ as
\begin{equation}
  \left[I,\rho_{n,m}\right]=(n-m)\rho_{n,m}.
\end{equation}

By construction the
Liouvillian preserves the number of excitations
and maps any $\rho_{n,m}$ into another
$\rho_{n,m}'$, i.e. it does not couple blocks.
Actually one can write an effective Liouville equation
for each subspace spanned by a pair of excitations $n$ and $m$ as
\begin{equation}
  \dot\rho_{n,m}={\mathcal L}_{n,m}\rho_{n,m},
\end{equation}
where each ${\mathcal L}_{n,m}$ is a superoperator acting on a $4-$dimensional space.
The full Liouvillian is simply the sum of all these terms and 
can be expressed as
\begin{equation}
  {\mathcal L}=\sum_{n,m}{\mathcal L}_{n,m}.
\end{equation}
If we express each block of the density matrix as a column vector
\begin{equation}
  \rho_{n,m}=
  \left(
  \begin{array}{cc}
  \rho_{n,m}^{1,1}&
  \rho_{n,m}^{1,0}\\ \\
  \rho_{n,m}^{0,1}&
  \rho_{n,m}^{0,0}
\end{array}
  \right)
  \equiv
  \left[
  \begin{array}{c}
  \rho_{n,m}^{1,1}\\
  \rho_{n,m}^{1,0}\\
  \rho_{n,m}^{0,1}\\
  \rho_{n,m}^{0,0}
\end{array}
  \right]
\end{equation}
then, the Liouville operator can be written in terms of the $4\times 4$ matrices 

\begin{eqnarray}\label{eq:Lmat}
 & {\mathcal L}_{n,m}=
\nonumber\\
 &\left[
\begin{array}{cccc}
 -\sqrt{mn} \gamma_0 & i g \sqrt{m} & -i g \sqrt{n} & \sqrt{mn} \gamma_1 \\
 i g \sqrt{m} & -2 i \delta -\frac{1}{2} \sqrt{mn} 
 \tilde\gamma & 0 & -i g \sqrt{n} \\
 -i g \sqrt{n} & 0 & 2 i \delta -\frac{1}{2} \sqrt{mn} 
 \tilde\gamma & i g \sqrt{m} \\
 \sqrt{mn} \gamma_0 & -i g \sqrt{n} & i g \sqrt{m} & -\sqrt{mn} \gamma_1
\end{array}
\right]
\end{eqnarray}
where we have defined
\begin{equation}
  \tilde\gamma=\gamma_1-\gamma_0.
\end{equation}
For $m=n$ each block reduces to the Liouville super-operator of a two-level atom
in thermal equilibrium with a reservoir and driven by a classical field.

The problem is now reduced to solving the eigenvalue problem for 
each non-hermitian $4\times 4$ matrix of the Liouvillian.
In this case, right and left eigenvectors are not the same and
the eigenvalue equations read
\begin{eqnarray}
  {\mathcal L}_{n,m}\hat \rho_{n,m}^{(j)}&=
  \lambda_{n,m}^{(j)}\hat\rho_{n,m}^{(j)}
  \nonumber\\
  \check \rho_{n,m}^{(j)}  {\mathcal L}_{n,m}&
  =\lambda_{n,m}^{(j)}\check\rho_{n,m}^{(j)}.
\end{eqnarray}
The biorthonormality condition is
\begin{equation}
  {\rm Tr}\left\{
  \check \rho_{n,m}^{(j)} 
  \,
  \hat \rho_{n',m'}^{(j')}
  \right\}=\delta_{j,j'}\delta_{n,n'}\delta_{m,m'}.
\end{equation}

The evolution of an initial density matrix $\rho(0)$ under a time independent Liouvillian reads\begin{equation}\label{eq:eveol}
  \rho(t)=e^{{\mathcal L} t}\rho(0).
\end{equation}
Using the spectral decomposition and biorthonormality this can be rewritten as
\begin{equation}\label{eq:spectral}
  \rho(t)=\sum_{n,m=0}^\infty
  \sum_{j=1}^4
  c^{(j)}_{n,m}e^{t \lambda_{n,m}^{(j)} }\hat\rho_{n,m}^{(j)}
\end{equation}
with the coefficients  
\begin{equation}
  c^{(j)}_{n,m}={\rm Tr}\left\{
  \check \rho^{(j)}_{n,m}\,\rho_0
  \right\}.
\end{equation}
Now that we have found the formal solution to the problem, we
can study the dynamics through the evaluation 
of relevant quantities. Here we shall concentrate
on the atomic system and consider its reduced density matrix,
which can be evaluated as
\begin{equation}
  \varrho=\sum_n \bra{n}\rho\ket{n}
\end{equation}
to obtain the $2\times 2$ matrix

\begin{eqnarray}\label{eq:redmat}
  \varrho=
  \left(
  \begin{array}{cc}
    \varrho_{11}&\varrho_{10}\\
    \\
    \varrho_{01}&\varrho_{00}
  \end{array}
  \right)
  =\sum_n 
  \left(
  \begin{array}{cc}
    \rho_{n,n}^{1,1}&\rho_{n+1,n}^{1,0}\\
    \\
    \rho_{n,n+1}^{0,1}&\rho_{n,n}^{0,0}
  \end{array}
  \right).
\end{eqnarray}
One can now evaluate the population inversion of the atomic system
\begin{equation}\label{eq:popinv}
  W(t)=\varrho_{11}(t)-\varrho_{00}(t)=2\varrho_{11}(t)-1
\end{equation}
as well as its purity
\begin{equation}\label{eq:pur}
  P(t)=\varrho_{11}^2(t)+\varrho_{00}^2(t)+2|\varrho_{01}(t)|^2.
\end{equation}

It is equally interesting to study the behaviour of the cavity. Then the 
Wigner or Husimi functions should be evaluated to get some insight of the action of the Kraus operators proposed by analyzing the corresponding phase space along the lines of ref.~\cite{plenio}. We leave this for future investigation.

\section{Explicit solutions for zero detuning}\label{sec:det0}
We now concentrate on the case
when the atom is in resonance with the mode, {\it i. e.} zero detuning
$\delta=0$. Here one is able to find simple explicit solutions
for eigenvalues and eigenvectors of the Liouville operator
of the master  \Eref{eq:master} by diagonalizing 
the blocks \eref{eq:Lmat}.
We shall rescale to unity coupling
between atom and cavity, that is $g=1$.
The four eigenvalues for each block are given by
\begin{eqnarray}\label{eq:eigval}
  \lambda_{n,m}^{(1)}&=-\frac{3\sqrt{m n}}{4}\tilde\gamma
  -\sqrt{
  \frac{m n }{16}\tilde\gamma^2
  - \left(\sqrt{m}+\sqrt{n}\right)^2}
  \nonumber\\
  \lambda_{n,m}^{(2)}&=-\frac{3\sqrt{m n}}{4}\tilde\gamma
  +\sqrt{
  \frac{m n }{16}\tilde\gamma^2
  - \left(\sqrt{m}+\sqrt{n}\right)^2}
  \nonumber\\
  \lambda_{n,m}^{(3)}&=-\frac{\sqrt{m n}}{4}\tilde\gamma
  -\sqrt{
  \frac{m n }{16}\tilde\gamma^2
  - \left(\sqrt{m}-\sqrt{n}\right)^2}
  \nonumber\\
  \lambda_{n,m}^{(4)}&=-\frac{\sqrt{m n}}{4}\tilde\gamma
  +\sqrt{
  \frac{m n }{16}\tilde\gamma^2
  - \left(\sqrt{m}-\sqrt{n}\right)^2}
\end{eqnarray}
Note that for $m=n$ the fourth eigenvalue is zero, which
corresponds to the stationary state of the dynamics. 

It is convenient to introduce
\begin{eqnarray}
  l_{n,m}^{(j)}&=\frac{\sqrt{m n}}{2}\tilde\gamma+\lambda_{m,m}^{(j)}
\end{eqnarray}
To simplify the notation in the following equations we shall 
write $\lambda_j=\lambda_{n,m}^{(j)}$ and 
$l_j=l_{n,m}^{(j)}$. With this notation, 
the full set of  left and right eigenvectors written in matrix form:
\begin{eqnarray}\label{eq:left}
  \check\rho_{n,m}^{1}&=
  \left(
  \begin{array}{cc}
  i\frac{(2+\gamma_0 l_1)(\sqrt m+\sqrt n)}
  {l_2(4+l_1\tilde\gamma)}
  & 
  -\frac{m\gamma_0+n\gamma_1-2l_1}
  {l_2(4+l_1\tilde\gamma)}
  \\
  \\
  \frac{n\gamma_0+m\gamma_1-2l_1}
  {l_2(4+l_1\tilde\gamma)}
  & 
  -i\frac{(2+\gamma_1 l_1)(\sqrt m+\sqrt n)}
  {l_2(4+l_1\tilde\gamma)}
  \end{array}
  \right)
  \nonumber\\
  \check\rho_{n,m}^{2}&=
  \left(
  \begin{array}{cc}
  -\frac{2+\gamma_0 l_2}
  {(4+l_2\tilde\gamma)}
  & 
  -i\frac{m\gamma_0+n\gamma_1-2l_2}
  {(\sqrt m+\sqrt n)(4+l_2\tilde\gamma)}
  \\
  \\
  i\frac{n\gamma_0+m\gamma_1-2l_2}
  {(\sqrt m+\sqrt n)(4+l_2\tilde\gamma)}
  & 
  i\frac{2+\gamma_1 l_2}
  {4+l_2\tilde\gamma}
  \end{array}
  \right)
  \nonumber\\
  \check\rho_{n,m}^{3}&=
  \left(
  \begin{array}{cc}
  i\frac{\sqrt m-\sqrt n}{\lambda_3-\lambda_4}&
  \frac{\lambda_3}{\lambda_3-\lambda_4}
  \\
  \\
  \frac{\lambda_3}{\lambda_3-\lambda_4} & 
  i\frac{\sqrt m-\sqrt n}{\lambda_3-\lambda_4}
  \end{array}
  \right)
  \nonumber\\
  \check\rho_{n,m}^{4}&=
  \left(
  \begin{array}{cc}
  \frac{\lambda_3}{\lambda_3-\lambda_4}
  &
  -i\frac{\sqrt m-\sqrt n}{\lambda_3-\lambda_4}
  \\
  \\
  -i\frac{\sqrt m-\sqrt n}{\lambda_3-\lambda_4}
  & 
  \frac{\lambda_3}{\lambda_3-\lambda_4} 
  \end{array}
  \right)
\end{eqnarray}

\begin{eqnarray}\label{eq:right}
  \hat\rho_{n,m}^{1}&=
  \left(
  \begin{array}{cc}
  i\frac{\sqrt m+\sqrt n}{l_2-l_1}&-\frac{l_2}{l_2-l_1}
  \\
  \\
  \frac{l_2}{l_2-l_1} & -i\frac{\sqrt m+\sqrt n}{l_2-l_1}
  \end{array}
  \right)
  \nonumber\\
  \hat\rho_{n,m}^{2}&=
  \left(
  \begin{array}{cc}
  -\frac{l_2}{l_2-l_1} & -i\frac{\sqrt m+\sqrt n}{l_2-l_1}
  \\
  \\
  i\frac{\sqrt m+\sqrt n}{l_2-l_1}& \frac{l_2}{l_2-l_1}
  \end{array}
  \right)
  \nonumber\\
  \hat\rho_{n,m}^{3}&=
  \left(
  \begin{array}{cc}
  i\frac{(2-\gamma_1 \lambda_4)(\sqrt m-\sqrt n)}
  {\lambda_3(4-\lambda_4\tilde\gamma)}
  & 
  -\frac{n\gamma_0+m\gamma_1+2\lambda_4}
  {\lambda_3(4-\lambda_4\tilde\gamma)}
  \\
  \\
  -\frac{m\gamma_0+n\gamma_1+2\lambda_4}
  {\lambda_3(4-\lambda_4\tilde\gamma)}
  & 
  i\frac{(2-\gamma_0 \lambda_4)(\sqrt m-\sqrt n)}
  {\lambda_3(4-\lambda_4\tilde\gamma)}
  \end{array}
  \right)
  \nonumber\\
  \hat\rho_{n,m}^{4}&=
  \left(
  \begin{array}{cc}
  \frac{2-\gamma_1 \lambda_3}
  {(4-\lambda_3\tilde\gamma)}
  & 
  i\frac{n\gamma_0+m\gamma_1+2\lambda_3}
  {(\sqrt m-\sqrt n)(4-\lambda_3\tilde\gamma)}
  \\
  \\
  i\frac{m\gamma_0+n\gamma_1+2\lambda_3}
  {(\sqrt m-\sqrt n)(4-\lambda_3\tilde\gamma)}
  & 
  \frac{2-\gamma_0 \lambda_3}
  {4-\lambda_3\tilde\gamma}
  \end{array}
  \right).
\end{eqnarray}
For  $m=n$ the proper limit
has to be taken in the fourth eigenvector and that is
\begin{eqnarray}
  \hat\rho_{n,n}^{4}&=
  \left(
  \begin{array}{cc}
  \frac{4+n\tilde\gamma\gamma_1}
  {8+n\tilde\gamma^2}
  & 
  -i\frac{2\sqrt n(\gamma_0-\gamma_1)}
  {8+n\tilde\gamma^2}
  \\
  \\
  i\frac{2\sqrt n(\gamma_0-\gamma_1)}
  {8+n\tilde\gamma^2}
  & 
  \frac{4+n\tilde\gamma\gamma_0}
  {8+n\tilde\gamma^2}
  \end{array}
  \right).
\end{eqnarray}

Having found explicit solutions for the
eigenvalue problem, we are going to use
them to investigate the dynamics of two simple
 initial conditions.

 \subsection{A dressed state as an initial condition}\label{sec:dress}
\begin{figure}[!t]
  \includegraphics[width=.49\textwidth]{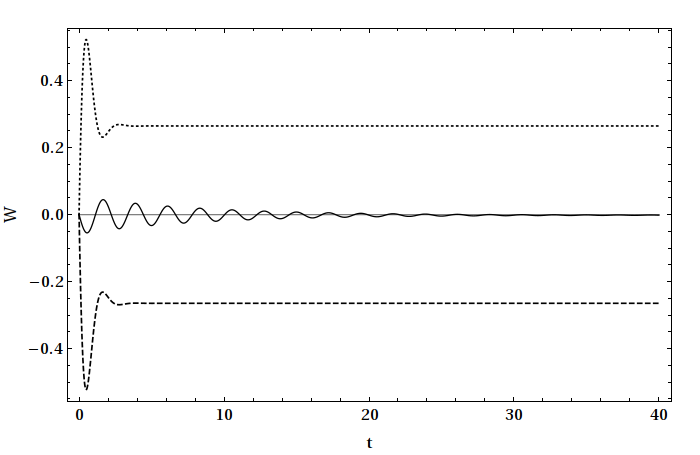}
  \includegraphics[width=.49\textwidth]{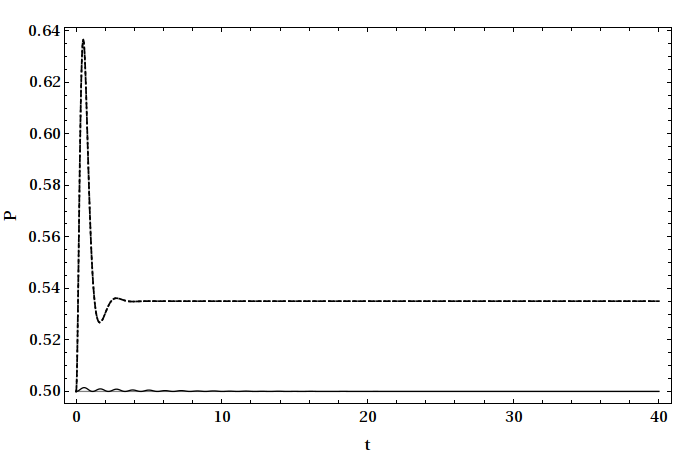}
  \caption{
  Population inversion $W$ (top) and purity $P$ (bottom) as a function
  of time for an initial dressed state. We show 4 curves,
  all with the same number of excitations $n=2$
  and $\gamma_0=\gamma_1$ for the gray curve,
  $\gamma_0=0.08$, $\gamma_1=0$ for the solid black
  curve, $\gamma_0=1.2$, $\gamma_1=0$ for the dashed curve
  and $\gamma_0=0$, $\gamma_1=1.2$ for the dotted curve.
  }\label{fig:dress}
\end{figure}

In this subsection we investigate the behaviour of an 
eigenvector of $H$, which is a stationary state
if the dissipation is turned off. In this case
with zero detuning ($\delta=0$) such state has the form
\begin{equation}
  \ket{\phi_n^+}=\frac{1}{\sqrt 2}\left(
  \ket{n-1,1}+\ket{n,0}
  \right).
\end{equation}
The eigenstates of $H$ have a definite number of excitations $n$,
and thus the  initial state can be expressed as 
\begin{eqnarray}\label{eq:dressinmat}
  \rho(0)=\rho_{n,n}(0)=
  \ketbra{\phi_n^+}{\phi_n^+}=
  \frac{1}{2}
  \left(
  \begin{array}{cc}
    1&1\\
    1&1
  \end{array}
  \right).
\end{eqnarray}
As different blocks do not couple we obtain
\begin{equation} 
  \rho(t)=\rho_{n,n}(t)=
  \sum_{j=1}^4{\rm Tr}\left\{
  \check\rho_{n,n}^{(j)}\ketbra{\phi_n^+}{\phi_n^+}
  \right\}e^{t\lambda_{n,n}^{(j)}}\hat\rho_{n,n}^{(j)}
\end{equation}
In this case we find that the relevant matrix elements are given by 
\begin{eqnarray}
  \rho_{nn}^{1,1}(t)&=
  \frac{2-\gamma_1\lambda_{n,n}^{(3)}}{4-\lambda_{n,n}^{(3)}\tilde\gamma}
  +
  \frac{\gamma_0-\gamma_1}{l_{n,n}^{(1)}-l_{n,n}^{(2)}}
  \sum_{j=1}^2
  \frac{l_{n,n}^{(j)^2}e^{\lambda_{n,n}^{(j)}t}}{8+2l_{n,n}^{(j)}\tilde\gamma}
  \nonumber\\
  \nonumber\\
  \rho_{n,n}^{1,2}(t)&=i\frac{\sqrt n(\gamma_0-\gamma_1)}{\tilde\gamma\lambda_{n,n}^{(3)}-4}
  -\tilde\gamma\frac{n e^{\lambda_{n,n}^{(3)}t}}{4\lambda_{n,n}^{(3)}}
  \nonumber\\
  &-
i\frac{\sqrt n(\gamma_0-\gamma_1)}
   {l_{n,n}^{(1)}-l_{n,n}^{(2)}}
  \sum_{j=1}^2
  \frac{l_{n,n}^{(j)}e^{\lambda_{n,n}^{(j)}t}}{4+l_{n,n}^{(j)}\tilde\gamma}
  (-1)^j
\end{eqnarray}
Once we have the density matrix, we can compute the reduced
density matrix for the atomic system from  
 \Eref{eq:redmat}. The off-diagonal terms
vanish and we are left with the diagonal
reduced density matrix for the atom
\begin{equation}\label{eq:dens00}
  \varrho=\left(
  \begin{array}{cc}
    \rho_{n,n}^{1,1} &0\\
    0& 1-\rho_{n,n}^{1,1}
  \end{array}
  \right).
\end{equation}
To visualize the dynamics we choose to evaluate the
population inversion $W(t)$ of the atomic state, that can be
obtained from  \Eref{eq:popinv}
and its purity $P(t)$ from  \Eref{eq:pur}.
This can be trivially achieved, noting that for the
reduced density matrix \eref{eq:dens00} the matrix element 
$\varrho_{11}=\rho_{n,n}^{1,1}$ gives us
all the information about the atomic subsystem.
In figure \ref{fig:dress} we have plotted these quantities, for different
choices of $\gamma_1$ and $\gamma_0$. 

First we will discuss the case $\gamma_0=\gamma_1$. This is plotted in
gray and the result are the constants values  $W=0$ and $P=1/2$ 
in figure \ref{fig:dress}. Of course this case includes the absence
of dissipation, where the dressed state is an eigenvector and therefore
 a stationary state. For non zero but equal dissipation constants, we 
 see that the state remains stationary.

For small dissipation constants where we set $\gamma_0=0.08$ and $\gamma_1=0$ (black curve)
we see small oscillations that eventually decay to the initial
values of $P$ and $W$.

Increasing in an asymmetric way the dissipation constants ($\gamma_0=0$, $\gamma_1=1.2$)  the oscillations
are strongly damped and  leading to a greater final population of the excited state (dotted line).
We find similar results
when ($\gamma_0=1.2$, $\gamma_1=0$), but here the ground state population
dominates for large times (dashed line).
For the last two conditions, the behaviour of purity is equivalent
and leads to a final steady state which is not a complete mixture as was the case for the initial state, i.e purity of the atom increases due to the external coupling.

\subsection{ A superposition  with  two  different number of excitations}\label{sec:sup}
As another example we now consider the
time evolution of an initial 
product state of the field and atom
subsystems, that is
\begin{equation}
  \ket{\Psi_0}= \cos\alpha\ket{n,0}+\sin\alpha\ket{n,1},
\end{equation}
The first  and second terms in the last expression
correspond, respectively, to states of $n$ and  $n+1$ excitations. 
Taking the outer product
of this vector we obtain the initial density matrix which
results in  a mixture of terms with these two excitations.
The resulting matrix can be expressed as
the sum of the four terms 
\begin{equation}
  \rho(t)=\rho_{n,n}(t)+\rho_{n,n+1}(t)
  +\rho_{n+1,n}(t)+\rho_{n+1,n+1}(t),
\end{equation}
and can be represented as a $4\times 4$ matrix.
Every single term can be computed using the spectral decomposition
as in \Eref{eq:spectral}. Taking partial trace
over the cavity's degree of freedom to obtain the reduced
density matrix of the atomic system, as in  \Eref{eq:redmat},
yields

\begin{eqnarray}
  \varrho(t)=
  \left(
  \begin{array}{cc}
    \rho_{n,n}^{1,1}(t)+\rho_{n+1,n+1}^{1,1}(t)&\rho_{n+1,n}^{1,0}(t)\\
    \\
    \rho_{n,n+1}^{0,1}(t)&\rho_{n,n}^{0,0}(t)+\rho_{n+1,n+1}^{0,0}(t)
  \end{array}
  \right).
\end{eqnarray}
The $2\times 2$ density matrix is determined by two of its 
elements as
\begin{eqnarray}
  \varrho_{11}(t)&=
  \cos^2\alpha\Bigg(
  \frac{4+n\tilde\gamma\gamma_1}{8+n\tilde\gamma^2}
  +\frac{l_{n,n}^{(1)}(2+l_{n,n}^{(1)}\gamma_1)
  e^{\lambda_{n,n}^{(1)}t}
  }{(4+l_{n,n}^{(1)}\tilde\gamma)(l_{n,n}^{(2)}-l_{n,n}^{(1)})}
  -\frac{l_{n,n}^{(2)}(2+l_{n,n}^{(2)}\gamma_1)
  e^{\lambda_{n,n}^{(2)}t}
  }{(4+l_{n,n}^{(2)}\tilde\gamma)(l_{n,n}^{(2)}-l_{n,n}^{(1)})}
  \Bigg)
  \nonumber\\
  &+\sin^2\alpha\Bigg(
  \frac{4+(n+1)\tilde\gamma\gamma_1}{8+(n+1)\tilde\gamma^2}
  -\frac{l_{n+1,n+1}^{(1)}(2+l_{n+1,n+1}^{(1)}\gamma_0)
  e^{\lambda_{n+1,n+1}^{(1)}t}
  }{(4+l_{n+1,n+1}^{(1)}\tilde\gamma)(l_{n+1,n+1}^{(2)}-l_{n+1,n+1}^{(1)})}
  \nonumber\\&
  +\frac{l_{n+1,n+1}^{(2)}(2+l_{n+1,n+1}^{(2)}\gamma_0)
  e^{\lambda_{n+1,n+1}^{(2)}t}
  }{(4+l_{n+1,n+1}^{(2)}\tilde\gamma)(l_{n+1,n+1}^{(2)}-l_{n+1,n+1}^{(1)})}
  \Bigg)
  \nonumber\\
  \varrho_{01}(t)&=
  \cos\alpha\sin\alpha
  \nonumber\\
  &\Bigg(
  \frac{(n\tilde\gamma+\gamma_1-2l_{n,n+1}^{(1)})
  e^{\lambda_{n,n+1}^{(1)}t}}{(4+l_{n,n+1}^{(1)}\tilde\gamma)(l_{n,n+1}^{(2)}-l_{n,n+1}^{(1)})}
  -
  \frac{(n\tilde\gamma+\gamma_1-2l_{n,n+1}^{(2)})e^{\lambda_{n,n+1}^{(2)}t}}{(4+l_{n,n+1}^{(2)}\tilde\gamma)(l_{n,n+1}^{(2)}-l_{n,n+1}^{(1)})}
  +
  \nonumber\\&
  \frac{(n\tilde\gamma+\gamma_0+2\lambda_{n,n+1}^{(4)})e^{\lambda_{n,n+1}^{(3)}t}}
  {(4-\lambda_{n,n+1}^{(4)}\tilde\gamma)(\lambda_{n,n+1}^{(4)}-\lambda_{n,n+1}^{(3)})}
  -
  \frac{(n\tilde\gamma+\gamma_0+2\lambda_{n,n+1}^{(3)})e^{\lambda_{n,n+1}^{(4)}t}}
  {(4-\lambda_{n,n+1}^{(3)}\tilde\gamma)(\lambda_{n,n+1}^{(4)}-\lambda_{n,n+1}^{(3)})}
  \Bigg)
\end{eqnarray}

\begin{figure}[!h]
  \includegraphics[width=.49\textwidth]{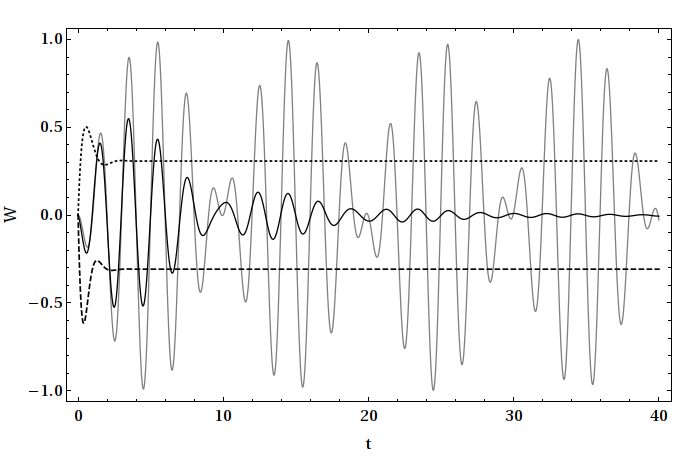}
  \includegraphics[width=.49\textwidth]{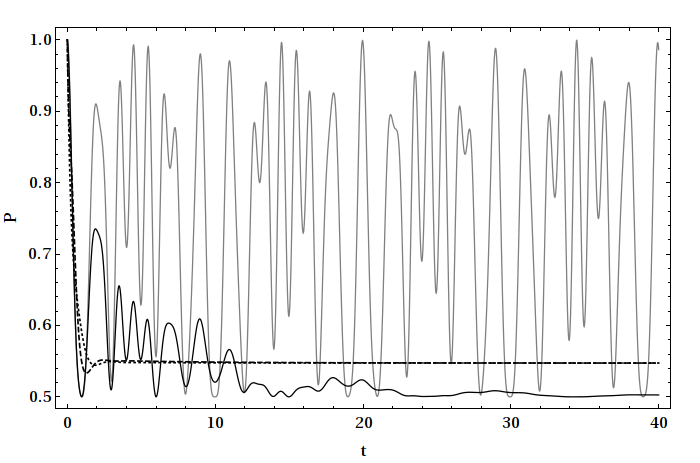}
  \caption{
  Population inversion $W$ (top) and purity $P$ (bottom) as a function
  of time for an initial dressed state. We show 4 curves,
  all with the same number of excitations $n=2$
  and $\gamma_0=\gamma_1=0$ for the gray curve,
  $\gamma_0=0.08$, $\gamma_1=0$ for the solid black
  curve, $\gamma_0=1.2$, $\gamma_1=0$ for the dashed curve
  and $\gamma_0=0$, $\gamma_1=1.2$ for the dotted curve.
  }\label{fig:sup}
\end{figure}

In the same way as in subsection \ref{sec:dress}, 
from the reduced density matrix of the atom, 
one can evaluate the inversion of population $W(t)$ \eref{eq:popinv}
and the purity $P(t)$ \eref{eq:pur}. In figure \ref{fig:sup} we have plotted
the corresponding values for different choices of $\gamma_1$ and $\gamma_0$.

For the gray curves ($\gamma_1=\gamma_0=0$) 
we can see that the initial state is not 
a stationary state of the system without dissipation.

Turning on one of the dissipation constants ($\gamma_0=0.08$, $\gamma_1=0$), 
the black curve shows
how the oscillations are damped to end up with a state with almost
equal population between excited
and ground state and close to the total mixture.

Increasing in an asymmetric way the dissipation constants results in a similar behaviour
as in the case where we used an initial dressed state. The oscillations
are strongly damped to end up with a greater occupation of the excited state
for for the dotted curve ($\gamma_0=0$, $\gamma_1=1.2$), and 
the opposite for the dashed curve ($\gamma_0=1.2$, $\gamma_1=0$).

\section{Conclusions}

We have displayed a new open system that can be solved analytically. From the point of view of solvable systems this is always an interesting step, both formally and also as a test ground for approximate or numerical solutions for problems that are not solvable analytically. Solvability here is based on conservation of the excitation number 
both in the unitary and the non-unitary evolution.

As far as the behaviour of the population inversion is concerned, we note that the two terms treat the cavity mode and spin asymmetrically if the two gammas are not equal.  Thus the observed result is not entirely surprising, but it may well deserve a more detailed study because purity will reach its lower limit only with symmetric dissipation .

While this model depends on the exact solvability of fourth order polynomials,
since Galois we know that such a general solution does not exist for higher degrees. 
We could try to cook up models that correspond to special solvable cases, but we would 
rather consider the following. If e.g. we have a three or more level atom 
(or a spin larger than $1/2$) or even say two two-level systems, according  to the 
discussion in the introduction the dimension of the matrix representation of the super
operator is the square of the dimension of the Hilbert space. Thus for spin 1 we would have
a $9\times 9$ 
matrix and for two spin $1/2$ particles a $16\times 16$ matrix form of the super-operator. Obviously these are not in general 
diagonalizable in closed form. Yet numerical diagonalization can be considered exact in 
the sense that arbitrary exactitude can be reached.   From that point on eigenvalues and the 
dual sets of eigenfunctions can be used to obtain the results required for a study of decoherence.

Other more complicated states, such as coherent states or thermal states can be used as initial states, and will be considered in future work.
The question how to construct a specific experiment that corresponds to these conditions, unfortunately, 
is still an open problem, but we are striving to find a solution.

\ack
J.M. Torres and T.H. Seligman would like to thank the Alexander von Humboldt foundation for support while working in this manuscript and Prof. W. Schleich for discussions and hospitality at the University of Ulm. The authors also acknowledge support
by the projects IN114310 by PAPIIT-UNAM 
and 79613 of CONACyT, Mexico.

\section*{References}
\bibliography{references}
\bibliographystyle{unsrt}

\end{document}